\documentclass[a4paper,11pt]{article}
\usepackage{jheppub} 

\usepackage{graphicx,multirow,subfigure}
\usepackage{float}
\usepackage{bm}
\usepackage{amsmath}
\usepackage{amssymb}
\usepackage{amscd}
\usepackage{latexsym}
\usepackage{slashed}
\usepackage{color}
\usepackage{graphicx}
\usepackage[normalem]{ulem}
\usepackage{color}
\definecolor{orange}{cmyk}{0,0.5,1,0}
\usepackage{verbatim}
\usepackage{rotating}
\usepackage{braket}
\usepackage{diagbox}
\newcommand{\mat}[1]{\begin{pmatrix} #1 \end{pmatrix}}

\def\lsim{\raise0.3ex\hbox{$\;<$\kern-0.75em\raise-1.1ex\hbox{$\sim\;$}}}
\def\gsim{\raise0.3ex\hbox{$\;>$\kern-0.75em\raise-1.1ex\hbox{$\sim\;$}}}

\usepackage{cancel}
\usepackage[utf8]{inputenc}
\newcommand{\snu}{\tilde{\nu}}
\newcommand{\snuRI}{\snu _1 ^{{\rm R,I}}}

\def\be{\begin{equation}}
\def\ee{\end{equation}}
\def\bea{\begin{eqnarray}}
\def\eea{\end{eqnarray}}

\title{Sneutrino Dark Matter in the  BLSSM}

\author[a,b]{Luigi Delle Rose,}
\author[c]{Shaaban Khalil,}
\author[a]{Simon J.D. King,}
\author[d]{Suchita Kulkarni,}
\author[e]{Carlo Marzo,}
\author[a,b]{Stefano Moretti}
\author[f]{and Cem S. Un}

\emailAdd{L.Delle-Rose@soton.ac.uk}
\emailAdd{Skhalil@zewailcity.edu.eg}
\emailAdd{SJD.King@soton.ac.uk}
\emailAdd{suchita.kulkarni@oeaw.ac.at}
\emailAdd{Carlo.Marzo@kbfi.ee}
\emailAdd{S.Moretti@soton.ac.uk}
\emailAdd{cemsalihun@uludag.edu.tr}

\affiliation[a]{School of Physics and Astronomy, University of Southampton, Highfield, Southampton SO17 1BJ, United Kingdom}
\affiliation[b]{Particle Physics Department, Rutherford Appleton Laboratory, Chilton, Didcot, Oxon OX11 0QX, United Kingdom}
\affiliation[c]{Center for Fundamental Physics, Zewail City of Science and Technology, Sheikh Zayed,12588 Giza, Egypt}
\affiliation[d]{Institut f{\"u}r Hochenergiephysik, {\"O}sterreichische Akademie der Wissenschaften, 
	Nikolsdorfer Gasse 18, 1050 Wien, Austria}
\affiliation[e]{National Institute of Chemical Physics and Biophysics, R{\"a}vala 10, 10143 Tallinn, Estonia}
\affiliation[f]{Department of Physics, Uluda\~{g} University, TR16059 Bursa, Turkey}

\abstract{In the framework of the $(B-L)$ Supersymmetric Standard Model (BLSSM), we assess the ability of ground and space based experiments to establish the nature of its prevalent Dark Matter (DM) candidate, the sneutrino, which could either be CP-even or -odd. Firstly, by benchmarking this theory construct against the results obtained by the Planck spacecraft, we extract the  portions of the BLSSM parameter space compliant with relic density data. Secondly, we show that, based on current sensitivities of the Fermi Large Area Telescope (FermiLAT) and their future projections, the study of high-energy $\gamma$-ray spectra will eventually enable us to extract evidence of this DM candidate through its annihilations into $W^+W^-$ pairs (in turn emitting  photons), in the form of both an integrated flux and a differential energy spectrum which cannot be reconciled with the assumption of DM being fermionic (like, e.g., a neutralino), although it should not be possible to distinguish between the scalar and pseudoscalar hypotheses. Thirdly, we show that, while underground direct detection experiments will have little scope in testing sneutrino DM, the Large Hadron Collider (LHC) may be able to do so in 
a variety of multi-lepton signatures, with and without accompanying jets (plus missing transverse energy), following data collection during Run 2 and 3. }

\begin{document} 
\maketitle
\flushbottom

\section{Introduction}
\label{sec:introduction}

In addition to solving major flaws of the Standard Model (SM), such as the  hierarchy problem, the absence of gauge coupling unification, etc., Supersymmetry   (SUSY) provides a candidate for Dark Matter (DM), when $R-$parity conservation is imposed at the low scale, which in turn requires the Lightest Supersymmetric Particle (LSP) to be stable. In fact, at present, the possibility of either scalar \cite{Cao:2017sju} or fermionic \cite{Choubey:2017yyn} DM is still allowed by  observations. Among the possible candidates, the weakly interactive massive particles (WIMP) are of special importance \cite{Arcadi:2017kky}, since they can potentially be observable in direct detection \cite{Akerib:2016lao,Aprile:2015uzo,Brink:2005ej,Tanaka:2011uf,Abbasi:2009uz}, indirect detection \cite{Atwood:2009ez,Abdallah:2016ygi} and collider experiments \cite{Buchmueller:2017qhf,Basalaev:2017hni,Chatrchyan:2012me}, in addition to the measurements of the WMAP \cite{Hinshaw:2012aka}  and Planck \cite{Ade:2015xua} satellites. Excluding the regions where a charged SUSY particle becomes LSP, the Minimal Supersymmetric SM (MSSM) has, in principle, two candidates for DM in the form of the lightest sneutrino or neutralino. The former is an example for a scalar DM, while the latter is a fermionic
example. However, due to its large interactions with the $Z$ boson, the Left-Handed (LH) LSP sneutrino case as a potential DM candidate 
 is excluded by direct LEP searches and cosmological observations \cite{Falk:1994es,Arina:2007tm}. Thus, it has been known for a long time that the neutralino LSP is the only available candidate to saturate the DM relic density in the MSSM. However, the latest observations from DM direct detection experiments  have brought about strict constraints  on the  neutralino composition. For example, the Higgsino-like and wino-like neutralino LSP yield large scattering cross sections with nuclei \cite{Hebbar:2017olk}, which are mostly excluded currently by the LUX experiment \cite{Akerib:2016lao}. Even though such a DM candidate can be allowed by all such constraints when it is made suitably heavy, this in turn means that the surviving candidate cannot be within the observability range of current experiments. Conversely, lower scattering cross sections with nuclei can be realised when the neutralino LSP is bino-like. However, in this case, its relic density is usually much larger than the latest results from the WMAP \cite{Hinshaw:2012aka} and Planck \cite{Ade:2015xua} satellites, except only in a small portion of the fundamental parameter space \cite{DelleRose:2017ukx}. 

Hence, quite apart from other motivations emerging from self-consistency of the SUSY theory (like the so-called $\mu$ problem) and other experimental data, the DM sector alone calls for some forms of extended SUSY scenario. Furthermore, 
neutrino mass generation remains a problem for minimal SUSY.  In contrast, a recent study \cite{DelleRose:2017ukx} has shown that a simple extension of the MSSM with a $(B-L)$ symmetry (BLSSM), i.e.,  $SU(3)_{C}\times SU(2)_{L}\times U(1)_{Y}\times U(1)_{B-L}$ with a gauged $U(1)_{B-L}$, enriches the variety of possible DM candidates  
 and can easily account for all experimental results mentioned above. Moreover, such an extension requires three Right-Handed
(RH) neutrinos and their SUSY partners to cancel the ensuing $U(1)_{B-L}$ anomalies. Therefore, the BLSSM is also a natural framework for implementing a seesaw mechanism which then provides a dynamics  for the generation of neutrino masses and mixings \cite{Wendell:2010md}. Herein, in addition to the neutralino LSP, the fundamental parameter space  also allows DM  solutions with  a sneutrino LSP able to saturate the DM relic density and to remain compliant with scattering processes against nuclei. However, to validate such a solution, one needs first to make sure that the sneutrino mass eigenstates should be formed mostly by the RH sneutrino, since the LH sneutrino is excluded as a DM candidate (as mentioned above). One of the challenges in finding such a sneutrino LSP as DM solution is  the constraint that the 125 GeV Higgs boson data yield a rather heavy mass spectrum for the SUSY particles, which in turn leads to heavy sneutrinos when a universal mass parameter is imposed for all scalars at the Grand Unification Theory (GUT) scale. In addition, $Z'$ and RH sneutrino masses receive contributions from the singlet Vacuum Expectation Values (VEVs), see later on, which are responsible for breaking the $U(1)_{B-L}$ symmetry, and the heavy mass bound on the $Z'$,  $M_{Z'} \geq 4$ \cite{ATLAS:2017wce}\footnote{However, a recent study has showed that this  bound  can be reduced  to about 3.6 TeV if ${\rm BR}(Z'\rightarrow l^+l^-) \lesssim 10\%$ \cite{Araz:2017wbp}.},  requires the singlet fields to develop large VEVs, which, again,  points to so heavy RH sneutrinos that can hardly be the LSP in the mass spectrum.
 
In this work, we investigate the feasibility of the RH  sneutrino LSP as a suitable DM candidate within the BLSSM framework, which embeds a  Type-I seesaw mechanism for the neutrino masses. In this case though, realising that consistency with relic density of  such a DM candidate is difficult due to the  tiny Yukawa couplings ($Y_{\nu} \lesssim 10^{-6}$) involved \cite{Abbas:2007ag}, one may be tempted to conclude that its observation would be difficult. This perception may be further reinforced by the fact, even though the RH sneutrino can interact with the $Z$ boson through the gauge kinetic mixing between $U(1)_{Y}$ and $U(1)_{B-L}$, such an interaction is strongly suppressed by the heavy mass bound on the gauge boson associated with the $(B-L)$ symmetry (the aforementioned $Z'$). These difficulties can however be overcome by identifying some new DM annihilation channels, which we will discuss below, 
 in which the specific $(B-L)$ sector plays a crucial role. In this case then, one may even attempt to extract evidence of such new DM dynamics which can be tested, if not at present, in near future experiments, both collider and astrophysical ones.  

The rest of the paper is organised as follows. We first start with discussing the BLSSM RH sneutrinos in Section \ref{sec:interactions} by studying their mass matrix and the fundamental parameters entering the calculation of observable quantities related to this potential DM state. Then Section \ref{sec:annihilation} presents the possible (co)annihilation channels and resulting relic abundance of the RH sneutrinos. Once consistent solutions are identified, we investigate possible signatures of RH sneutrino DM in Fermi Large Area Telescope (FermiLAT)  and 
Large Hadron Collider (LHC) data  for some benchmark points in Sections \ref{sec:InDD} and \ref{sec:LHC}, respectively. 
Finally, we conclude in Section \ref{sec:conclusion}.

%%%%%%%%%%%%%%%%%%%%%%%%%%%%%%%%%%
\section{RH Sneutrinos in the BLSSM}
\label{sec:interactions}
\begin{figure}[th]
\centering
\includegraphics[width=.8\linewidth]{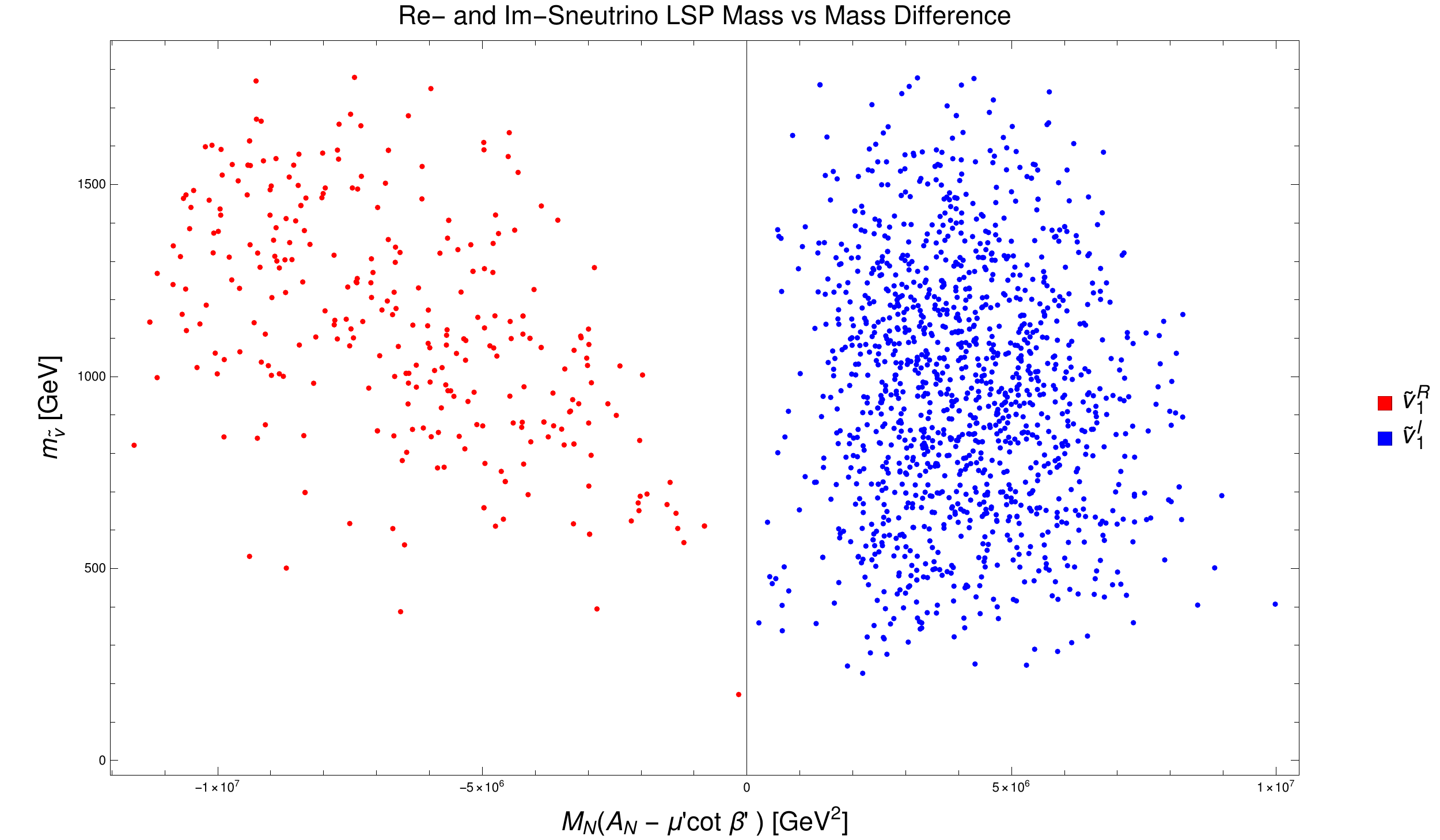}
\caption{Masses of real and imaginary RH sneutrino LSP candidates are plotted against the mass difference of the two eigenstates, $M_N(\mu'\cot\beta' - A_N)$.}
\label{fig:MSneu_vs_Mass_Diff}
\end{figure}

\noindent
We now consider the RH sneutrino sector in the BLSSM model. With a TeV scale
BLSSM with Type-I seesaw and very small neutrino Yukawa coupling, $Y_\nu \lsim {\cal O}(10^{-6})$, the sneutrino mass matrix, in the basis ($\tilde{\nu}_L,\tilde{\nu}_L^\ast,\tilde{\nu}_R,\tilde{\nu}_R^\ast$), is approximately given by a $2\times 2$ block diagonal matrix, where the element $11$ of this matrix is given by the diagonal LH sneutrino mass matrix and the element $22$ represents the RH sneutrino mass matrix, $M_{RR}$, defined as \cite{OLeary:2011vlq}  
	\begin{eqnarray}
		M^2_{RR} & = & \mat{M_N^2 + m_{\tilde{N}}^2 + m_D^2+ \frac{1}{2}M_{Z'}^2\cos 2\beta' & M_N(A_N-\mu'\cot\beta')\\
		M_N(A_N-\mu'\cot\beta') & M_N^2+m_{\tilde{N}}^2+m_D^2+\frac{1}{2}M_{Z'}^2\cos 2\beta'}.
	\end{eqnarray}
It is notable that a large mixing between the RH sneutrinos and RH antisneutrinos is quite plausible, since it is given in terms of large Yukawa couplings, $Y_N \sim {\cal O}(1)$.  Therefore, $\tilde{\nu}_R,\tilde{\nu}_R^\ast$ are not the mass eigenstates. The mass splitting and mixing between the RH sneutrino $\tilde{\nu}_R$ and RH antisneutrino $\tilde{\nu}_R^*$ are a result of the induced $\Delta L =2$ lepton number violating term $M_N N^cN^c$. One can show that the mass eigenvalues of RH sneutrinos are given by \cite{Elsayed:2012ec,Khalil:2011tb}
\begin{equation}
m^2_{\tilde{\nu}_{\mp}} = M_N^2+m_{\tilde{N}}^2+m_D^2+\frac{1}{2}M_{Z'}^2\cos 2\beta'  \mp  \Delta m_{\tilde{\nu}_R}^2,
\label{eq:mass_splitting}
\end{equation}
where $\Delta m_{\tilde{\nu}_R}^2 = \Big\vert M_N( A_N - \mu'\cot\beta') \Big\vert$ and the mass eigenstates $\tilde{\nu}_{\mp}$ are defined in terms of $\tilde{\nu}_R,\tilde{\nu}_R^\ast$ as follows:
\bea 
\tilde{\nu}_{-} &=& \frac{-i}{2}  \left(  e^{i \phi /2} \tilde{\nu}_R - e^{-i \phi /2} \tilde{\nu}_R^\ast \right) ,\\
\tilde{\nu}_{+} &=& \frac{1}{2}  \left( e^{i \phi /2} \tilde{\nu}_R + e^{-i \phi /2} \tilde{\nu}_R^\ast \right), 
\eea 
where $\phi$ is the phase of the off-diagonal element of $M_{RR}$, {i.e.}, $\phi = {\rm arg}( M_N(A_N - \mu'\cot\beta'))$. In case of real soft SUSY breaking terms, one finds $\phi=0$ or  $\phi=\pi$, depending on the relative sign of $A_N$ and $\mu'$. In the former case, we see that $\tilde{\nu}_{-} (\phi =0)={\rm I}(\tilde{\nu}_R) \equiv \tilde{\nu}_{_1}^{\rm I}$, so the lightest state is an imaginary sneutrino with $m_{  \tilde{\nu}_{_1}^{\rm I}}= m_{\tilde{\nu}_{-}}$  and the real type, ${\rm R}(\tilde{\nu}_R) \equiv \tilde{\nu}_1^{\rm R}$, has a larger mass $m_{  \tilde{\nu}_1^{\rm R}}= m_{\tilde{\nu}_{+}}$. The other possibility is $\phi=\pi$, where now $\tilde{\nu}_{-} (\phi =\pi)= \tilde{\nu}_{_1}^{\rm R}$ is the lightest state with $m_{  \tilde{\nu}_{_1}^{\rm R}}= m_{\tilde{\nu}_{-}}$  and $\tilde{\nu}_R^{\rm I}$ is heavier with $m_{  \tilde{\nu}_R^{\rm I}}= m_{\tilde{\nu}_{+}}$. One can see the behaviour in Fig. \ref{fig:MSneu_vs_Mass_Diff}. When the mass difference is positive, $\phi=0$ and so the $\tilde{\nu}_{_1}^{\rm I}$ acquires the lightest mass  $m_{\tilde{\nu}_{-}}$. In the case of a negative mass difference ($\phi =\pi$), one has a $\tilde{\nu}^{\rm R}_{_1}$ LSP, with mass $m_{\tilde{\nu}_{-}}$. In general, one finds that $M_N(A_N - \mu'\cot\beta')$ tends to be positive and so there are many more CP-odd sneutrino LSPs than CP-even ones (by a factor of $\sim 10$).

\begin{figure}[t]
	\begin{center}
	\subfigure{\includegraphics[scale=0.75]{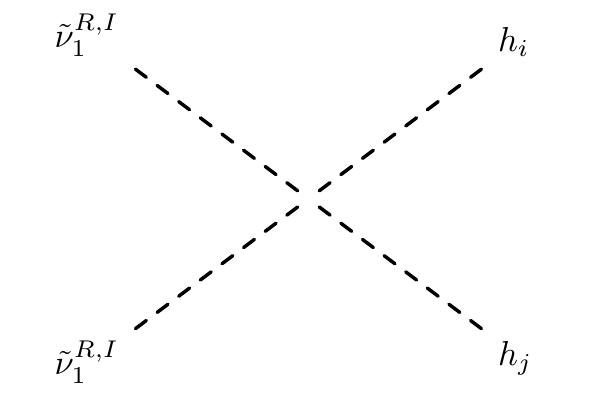}}\hspace{0.5cm}
	\subfigure{\includegraphics[scale=0.75]{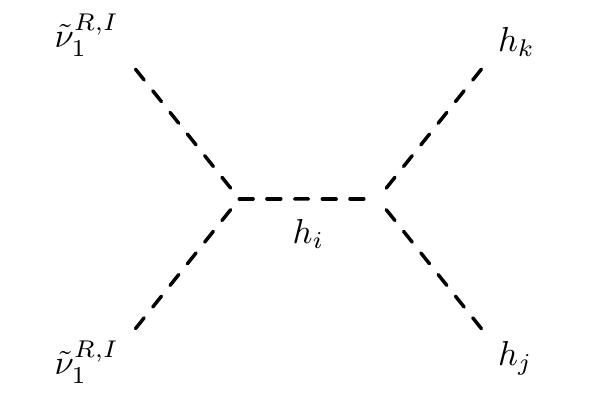}}\hspace{0.5cm}
	\subfigure{\includegraphics[scale=0.75]{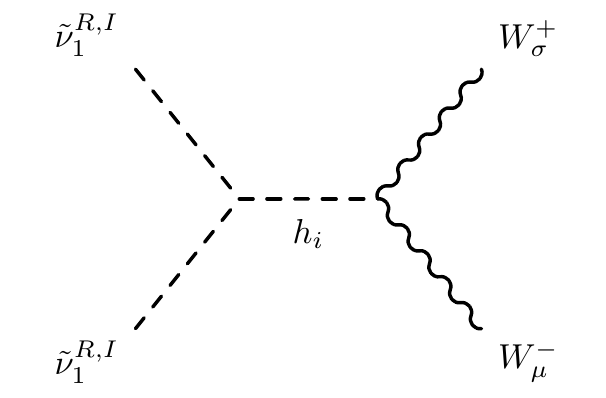}}
	\caption{Feynman diagrams of the dominant interaction terms of two real or two imaginary RH sneutrinos.}
	\label{fig:feyn_diag_sneutrinos}
\end{center}
\end{figure}

Now, we briefly describe the relevant interactions of sneutrino DM, for $\tilde{\nu}_R^{\rm I}$ and $ \tilde{\nu}_R^{\rm R}$ LSPs. The relic abundance of the sneutrino DM is a direct consequence of the strength of these interactions, in addition to revealing what signatures this DM candidate may provide. The main interactions which contribute to the annihilations of the sneutrino DM are given by  four-point interaction $\left( \tilde{\nu}^{({\rm R,I})}_{{1}} \tilde{\nu}^{({\rm R,I})}_{{1}} \rightarrow h_i h_j \right)$ and processes mediated by the CP-even Higgs sector $\left(\tilde{\nu}^{({\rm R,I})}_{{1}} \tilde{\nu}^{({\rm R,I})}_{{1}} \rightarrow  h_i \rightarrow h_i h_j~{\rm or} ~W^+ W^- \right) $, as shown in Fig.~\ref{fig:feyn_diag_sneutrinos}.
With $Y_\nu \ll 1$, the Lagrangian of these interactions can be written as follows:
{\small \begin{eqnarray}
	\mathcal{L} &\supset i \Biggl\{ \left(\tilde{\nu}^{{\rm R,I}}_{{1}} \right)^2 h_i \sum_{a=1}^{3}(Z^{({\rm R,I})}_{1 3 + a})^2  \Big[\frac{g_{B}^{2}}{2} \Big( v_{\eta} Z_{{i 3}}^{H} - v_{\bar{\eta}} Z_{{i 4}}^{H} \Big) \pm \sqrt{2} \Big( Z_{{i 4}}^{H} \mu _\eta Y_{x,{a a}} - Z_{{i 3}}^{H}T_{x,{a a}}\Big) -4v_\eta Z_{{i 3}}^{H} Y_{x,{a a}} ^2 \Big] \nonumber\\
	&+\left(\tilde{\nu}^{{\rm R,I}}_{{1}}\right)^2 h_i h_j \sum_{a=1}^{3}(Z^{({\rm R,I})}_{1 3 + a})^2  \Big[\frac{g_{B}^{2}}{2} \Big(  Z_{{i 3}}^{H} Z_{{j 3}}^{H} -  Z_{{i 4}}^{H} Z_{{j 4}}^{H} \Big) + \frac{g_{B} g_{YB}}{4} \Big( Z_{{i 1}}^{H} Z_{{j 1}}^{H} - Z_{{i 2}}^{H} Z_{{j 2}}^{H} \Big) - 4 Z_{{i 3}}^{H} Z_{{j 3}}^{H}Y_{x,{a a}} ^2 \Big] \nonumber\\
	&+ \left( h_i h_j h_k\right) g_{B}^{2} \Big[ v_{\eta}\Big(-   3  Z_{{i 3}}^{H} Z_{{j 3}}^{H} Z_{{k 3}}^{H} +     Z_{{i 3}}^{H} Z_{{j 4}}^{H} Z_{{k 4}}^{H} + Z_{{i 4}}^{H} Z_{{j 3}}^{H} Z_{{k 4}}^{H}+  Z_{{i 4}}^{H} Z_{{j 4}}^{H} Z_{{k 3}}^{H}   \Big) \nonumber \\
	&\hspace{2.12cm}+v_{\bar{\eta}} \Big(  Z_{{i 3}}^{H} Z_{{j 3}}^{H} Z_{{k 4}}^{H} +   Z_{{i 3}}^{H} Z_{{j 4}}^{H} Z_{{k 3}}^{H} + Z_{{i 4}}^{H} Z_{{j 3}}^{H} Z_{{k 3}}^{H}      -3  Z_{{i 4}}^{H} Z_{{j 4}}^{H} Z_{{k 4}}^{H}      \Big) \Big] \nonumber\\  
	&+h_i W^- _\mu W^+ _\sigma \frac{g_{2}^{2}}{2}  \Big(v_d Z_{{i 1}}^{H}  + v_u Z_{{i 2}}^{H} \Big)\Big(g^{\sigma \mu}\Big) \Biggr\},
	\label{eq:int_lagr}
	\end{eqnarray}}
where $h_i$ is one of the four mixed CP-even Higgs mass eigenstates \cite{Abdallah:2014fra} ($h_1$ is the lightest SM-like Higgs, $h_2$ is the light $(B-L)$-like Higgs, $h_3$ is the heavy MSSM-like Higgs and $h_4$ is the heavy $(B-L)$-like state). These states are all mixed and the matrix which diagonalises the Higgs mass matrix is written as $Z^H$. There are four Higgs VEVs, corresponding to the MSSM $H_u$ and $H_d$ doublets and the BLSSM $\eta$ and $\bar{\eta}$ singlets, written as $\left( v_u,~v_d,~v_\eta,~v_{\bar{\eta}} \right)$, respectively. The diagonalising mass matrices for the CP-even and CP-odd sneutrinos are denoted by $Z^{({\rm R,I})}$ while the $Y_{x, aa}$'s are the Yukawa couplings for the RH neutrinos, which are assumed to be diagonal along with the trilinear couplings, the $T_{x, aa}$'s. The gauge couplings $g_{B}$ and $g_{YB}$ will be rotated, along with the (unseen) $g_{YY}$ and $g_{BY}$ couplings, to become the physical $g_1$, $g_{BL}$ and $\tilde{g}$ couplings. 

%%%%%%%%%%%%%%%%%%%%%%%%%%%%%%%
%
\section{Annihilation Cross Section and DM Relic Abundance}
\label{sec:annihilation}
\begin{figure}[t]
\centering
\includegraphics[width=12cm, height=6.5cm]{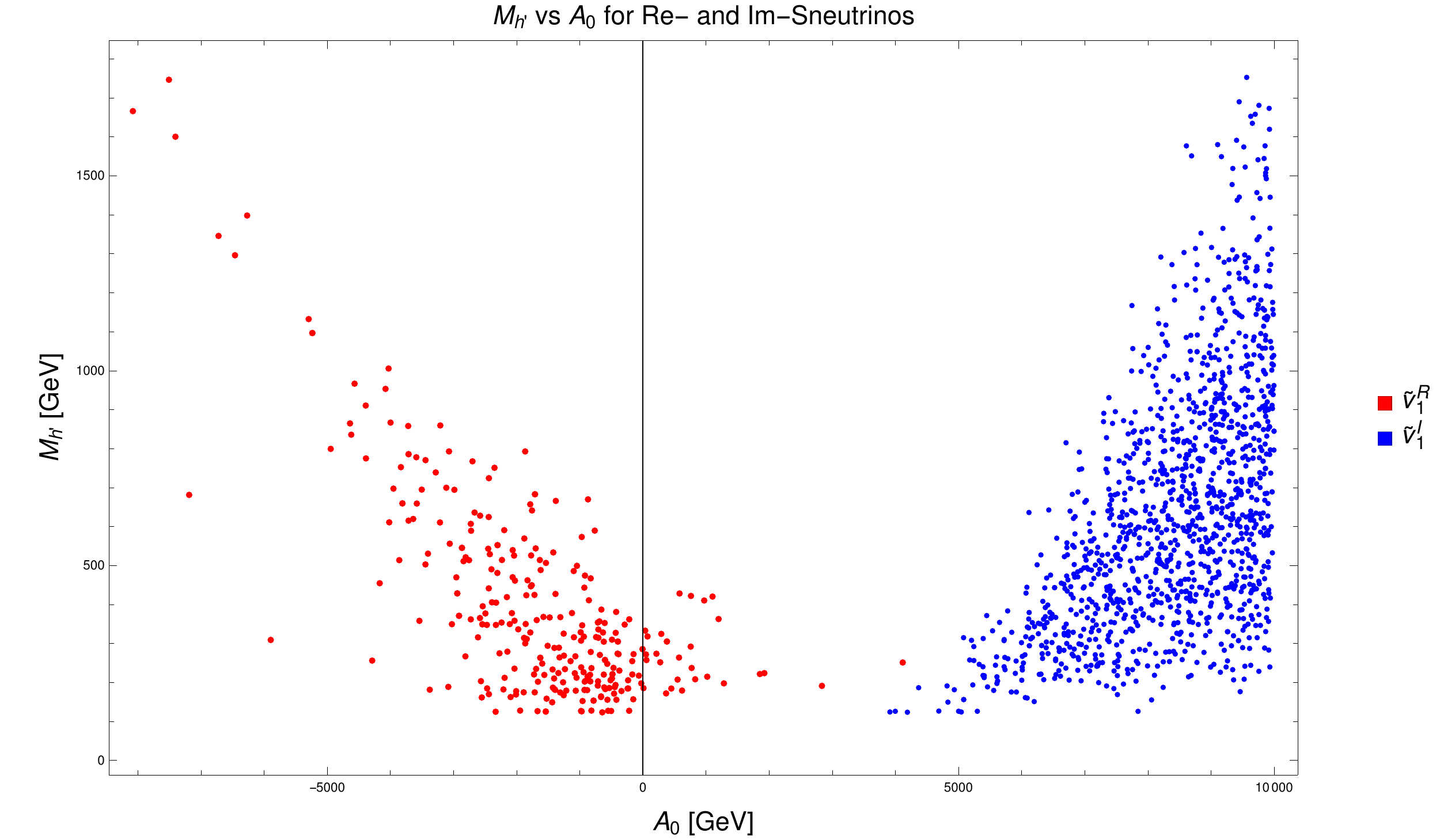}
\caption{Mass of lightest $(B-L)$-like Higgs versus the GUT parameter $A_0$, for CP-even sneutrino LSPs (red) and CP-odd sneutrino LSPs (blue).}
\label{fig:mh2_vs_a0_Re_Im}
\end{figure}

The two CP-eigenstate RH sneutrinos, $\tilde{\nu}_1^{\rm I}$ and $\tilde{\nu}_1^{\rm R}$, produce different phenomena in respect of the cross sections of their annihilation channels, which may yield detectable consequences in cosmological measurements. The DM is  annihilating at low (thermal) energies, so the final product masses must be $\lesssim 2M_{\snu}$. As indicated by the interaction terms in (\ref{eq:int_lagr}), the highest cross section channel (for both CP-even and -odd) is $\snu \snu \rightarrow h' h'$\footnote{For ease of notation, hereafter, we identify $h'\equiv h_2$.}, as long as $M_{h'}<M_{\snu}$. If this is not the case then the next highest cross section channel is $\snu \snu \rightarrow W^+ W^-$. We find that other channels have small contributions to the total annihilation cross section in comparison to these two. So, what separates the phenomena of the real and imaginary sneutrinos is then simply the mass relation between $h'$ and $\snu$. If $M_{\snu} > M_{h'}$, the annihilation cross section will be dominated by the $h'h'$ production and, if not, then $W^+ W^-$. In order to determine which mass is larger, and hence the phenomenology of a given state, we must consider the dependence of the mass splitting relation (\ref{eq:mass_splitting}) on the trilinear coupling $A_0$. This initial input parameter will determine the properties of our sneutrino LSP at the low scale. For $A_0 \lesssim 0$, this mass splitting will favour a lower mass CP-even sneutrino and hence LSP, while  for $A_0 \gtrsim 0$ we find CP-odd LSPs. The exact details are discussed previously, in Section \ref{sec:interactions}, but one finds this general trend, as seen in Fig.  \ref{fig:mh2_vs_a0_Re_Im}. 

Now, we turn to how the lightest $(B-L)$ Higgs, $h'$, which  is affected by $A_0$ as follows:
\begin{equation}
M_{h'} = \frac{1}{2}\Big[ (m_{A'}^2 + M_{Z'}^2) - \sqrt{m_{A'}^2 + M_{Z'}^2) ^2 - 4m_{A'}^2 M_{Z'}^2 \cos ^2 2 \beta '} ~\Big],
\end{equation}
where $m_{A'}^2$ is the mass of the $(B-L)$ CP-odd Higgs. It is this CP-odd mass which is affected by the trilinear coupling $A_0$ which causes $M_{h'}$'s dependence. Fig. \ref{fig:mh2_vs_a0_Re_Im} displays this relation and we see that, for large positive $A_0$ values, a wide range of $M_{h'}$ masses are allowed ($\sim 100 - 2000$ GeV) whereas, for $A_0 \lesssim 0$, lower $M_{h'}$ values are favoured, with the largest density of points over the interval $\sim 100 - 500$ GeV.

\begin{figure}[t]
%\begin{tabular}{c c}
\begin{center}
\includegraphics[width=7cm, height=6cm]{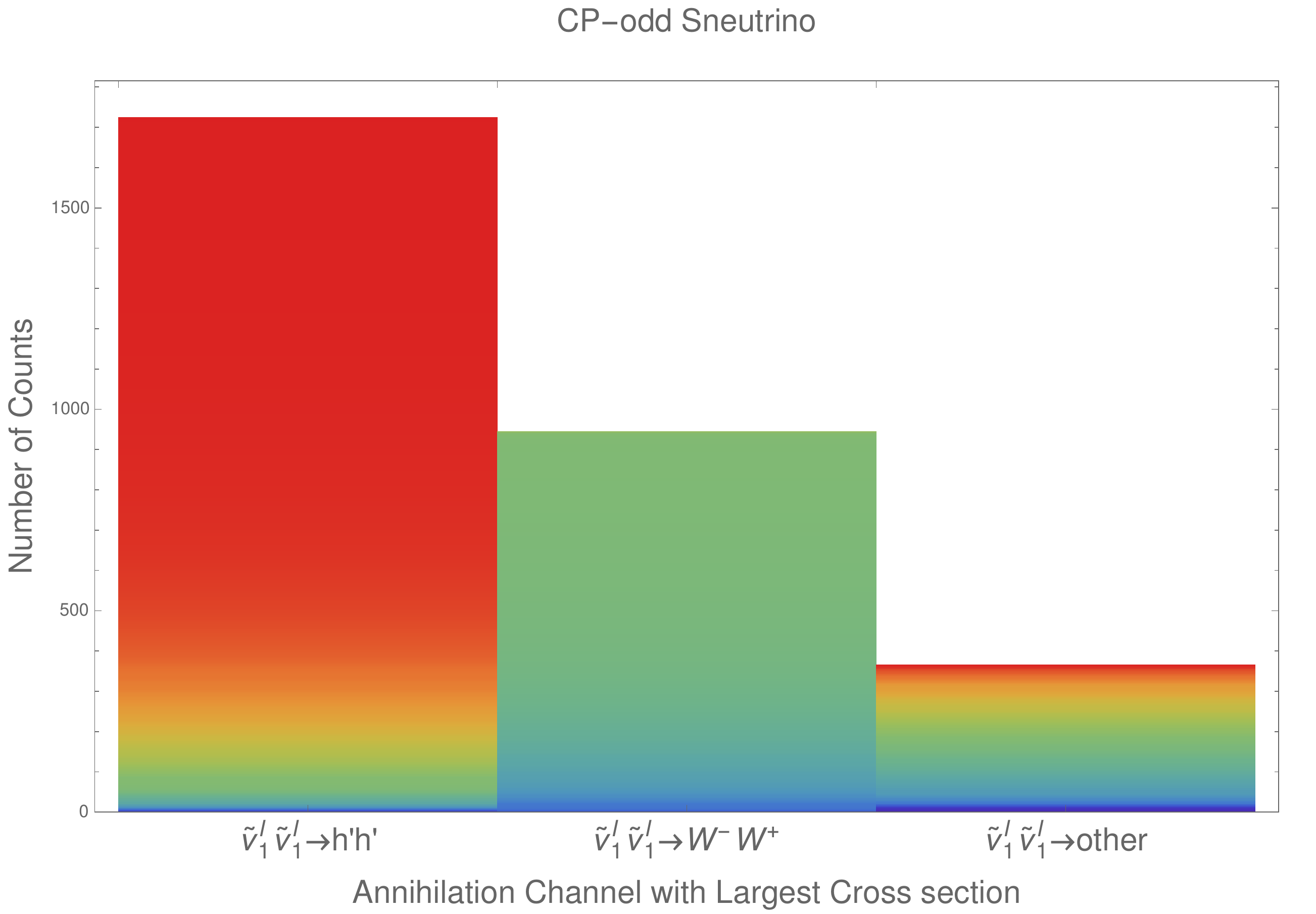} ~~~~ \includegraphics[width=7cm, height=6cm]{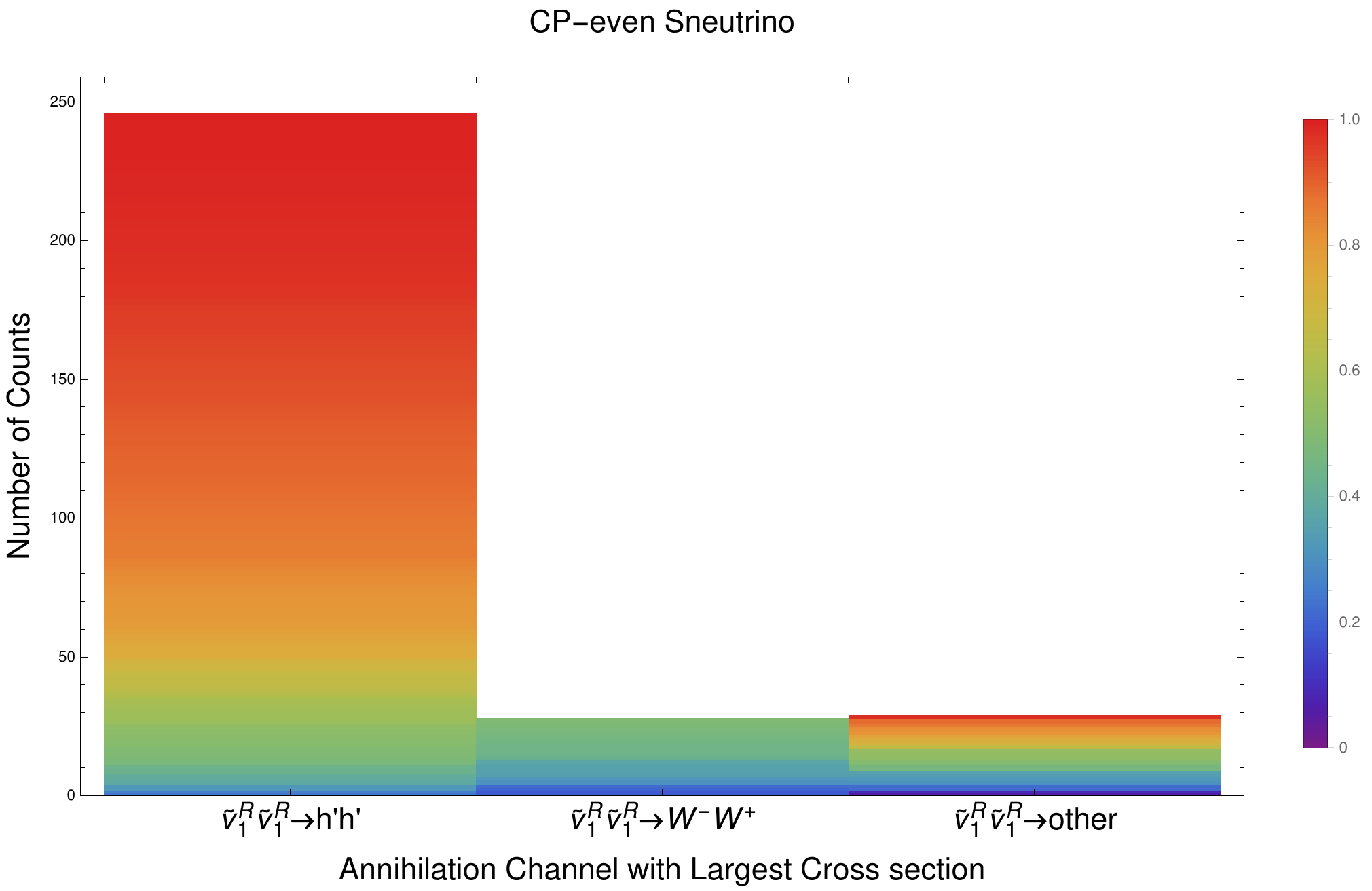}
%\end{tabular}
\end{center}
\caption{Histogram counting the number of spectrum points with the largest annihilation cross section being in either the $h'h'$, $W^+ W^-$ or other channel. This has been done for spectrum points which have a CP-odd (left) or CP-even (right) sneutrino LSP. Each count is also coloured by the normalised cross section (so that the sum of annihilation cross section channels for a given point is unity), where a red coloured point means the given annihilation channel has a  larger cross section.}
\label{fig:Heat_Map_Histogram}
\end{figure}

Combining this trend with larger mass scales for CP-even sneutrinos, as seen in Fig. \ref{fig:MSneu_vs_Mass_Diff}, provides us with two general cases based on the GUT parameters. Firstly, $A_0$ is negative, the sneutrino LSP is CP-even, with $m_{\snu} \gtrsim 500$ GeV and $M_{h'} \lesssim 500$ GeV, hence, in general, $m_{\snu} > M_{h'}$. The other possibility is that $A_0$ is positive, here, the sneutrino LSP is CP-odd and both masses are similar, $100\lesssim m_{\snu},M_{h'} \lesssim 2000$ GeV. Further, there are cases where $m_{\snu}$ is larger {and} also $M_{h'}$ is larger. This behaviour is reflected in Fig.  \ref{fig:Heat_Map_Histogram}, where the histogram counts the number of spectrum points where the annihilation channels $h'h'$, $W^- W^+$ or something else have the largest cross section. The different spectrum points are coloured according to the value of their normalised annihilation cross section for a particular channel (e.g., $\sigma(\snu \snu \rightarrow h' h')/\sigma(\snu \snu \rightarrow X)$, for any combination of particles $X$). One can see that the $h'h'$ decay has in general a very large cross section, in comparison to $W^- W^+$. For the CP-odd case, the majority of spectrum points have $m_{\snu} > M_{h'}$, but there are still a considerable number of solutions for which this is not the case,  so one would expect to see $W^- W^+$- annihilations for these. We also note that the cross section for CP-odd annihilations into $h'h'$  is larger than that for CP-even annihilations into $h'h'$, ({i.e.}, more red points in the former channel than for the latter onel). However, for the CP-even case, a much smaller region of parameter space allows for these $W^- W^+$ decays. Notice that, as it will be discussed in the next section, these charged decay products may account for an observable $\gamma$-ray spectrum.

These annihilation cross sections will be what determine the relic abundance of the sneutrinos. In this work we consider a standard cosmological scenario, where the DM particles were in thermal equilibrium with the SM ones  in the early Universe and decoupled when the temperature fell below their relativistic energy. The relic density of our sneutrino species is written as \cite{Lee:2007mt}:
\be
\Omega h^2 _{\snu _1 ^{{\rm R,I}}} = \frac{2.1 \times 10^{-27} \rm{cm}^3 \rm{s}^{-1}}{\braket{\sigma_{\snu _1 ^{{\rm R,I}}}^{\rm{ann}}v}} \left( \frac{x_F}{20}\right) \left(\frac{100}{g_* (T_F)}\right) ^{\frac{1}{2}},
\ee
where $\braket{\sigma_{\snu _1 ^{{\rm R,I}}}^{\rm{ann}}v}$ is a thermal average for the total cross section of annihilation to SM objects multiplied by the relative sneutrino velocity, $T_F$ is the freeze out temperature, $x_F \equiv m_{\snuRI}/T_F \simeq \mathcal{O}(20)$ and $g_*(T_F) \simeq \mathcal{O}(100)$ is the number of degrees of freedom at freeze-out.

\begin{figure}[h]
	\centering
	\includegraphics[width=10cm, height=6cm]{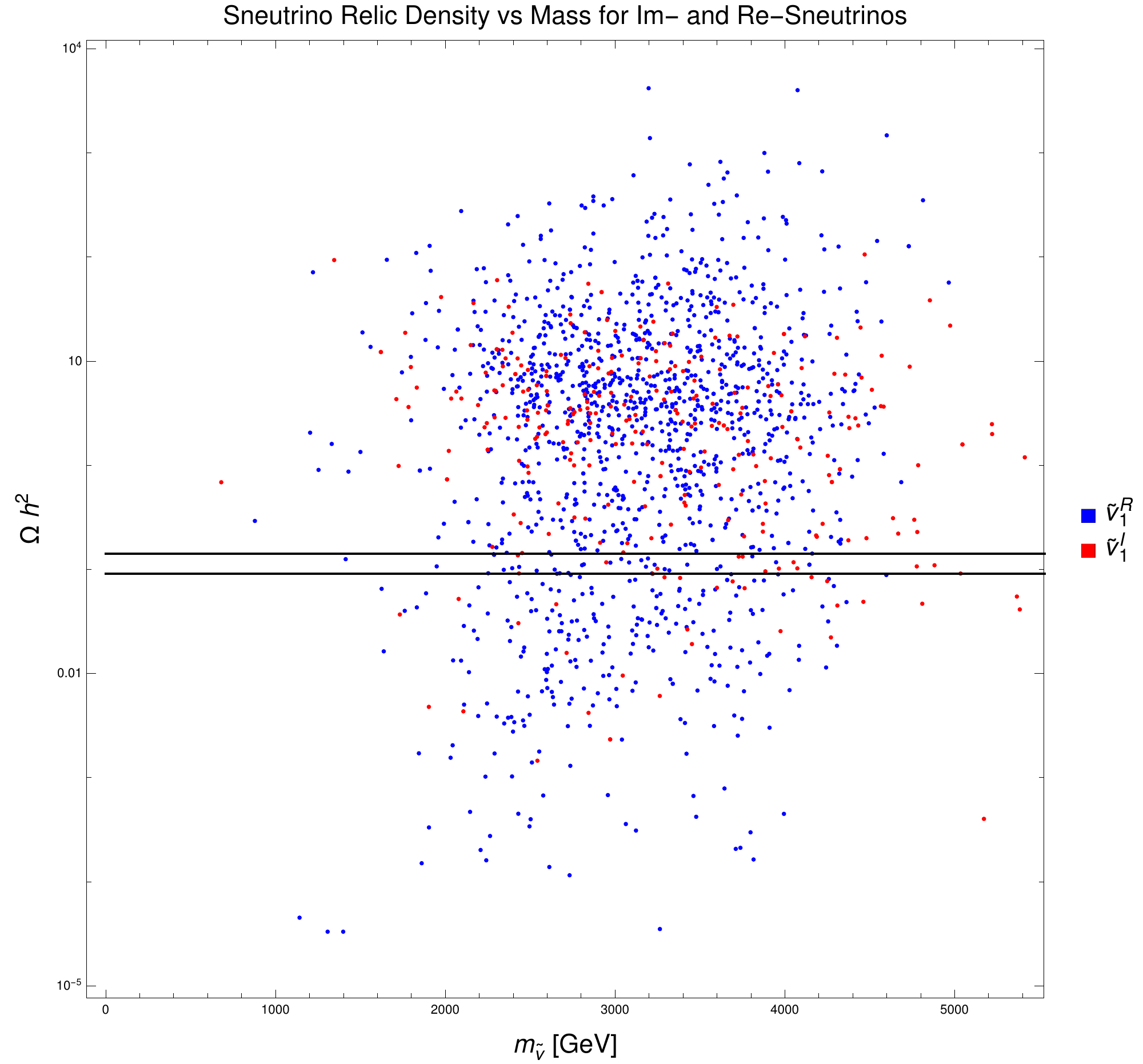}
	\caption{Relic density of CP-even and CP-odd sneutrinos versus their mass in GeV, where horizontal lines correspond to the Planck limits for the relic abundance.}
	\label{fig:Sneutrino_Relic}
\end{figure}

Fig. \ref{fig:Sneutrino_Relic} shows the thermal relic abundance for sneutrinos. This has been computed by micrOMEGAs  \cite{Belanger:2006is,Belanger:2013oya} and one can see that both CP-even and CP-odd candidates are allowed by current limits of $0.09 < \Omega h^2 < 0.14$, which is the $2\sigma$ allowed region by the Planck collaboration \cite{Ade:2015xua}. These points also satisfy the HiggsSignals/HiggsBounds
\cite{Bechtle:2013xfa,Bechtle:2013wla} constraints (that the lightest CP-even Higgs must be SM-like and subject to negative Higgs searches), in addition to SUSY mass bounds for gluinos, staus, neutralinos, charginos and stops \cite{Patrignani:2016xqp}. 

\section{Indirect Detection}
\label{sec:InDD}

\begin{figure}[t]
\centering
\includegraphics[width=12cm, height=7.5cm]{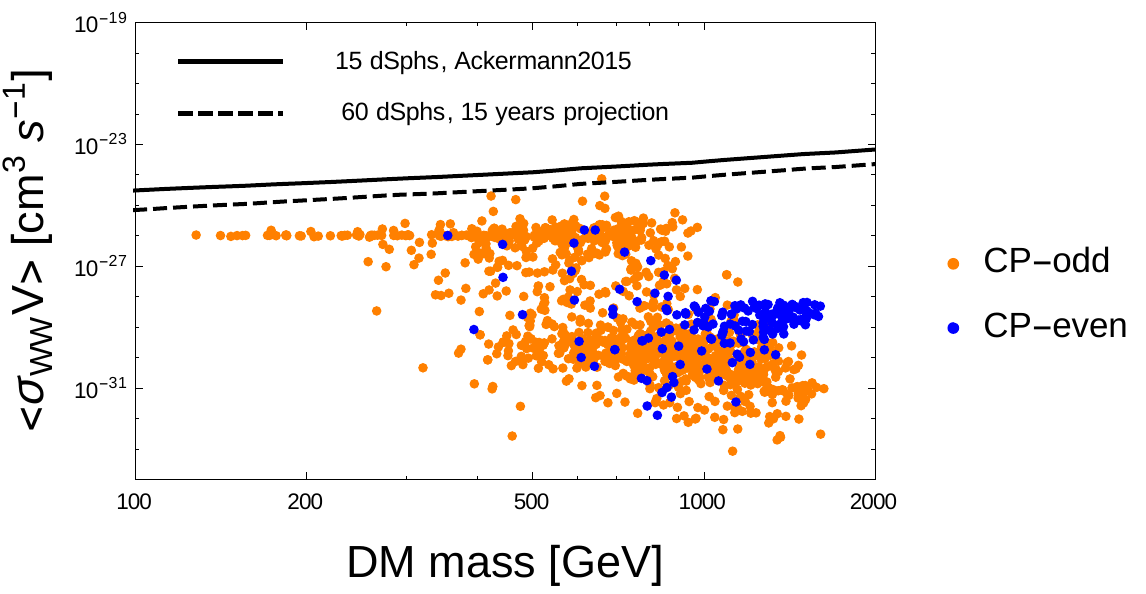}
\caption{Thermal cross section for DM DM $\to W^+W^-$ annihilation as predicted by theory as a function of the DM mass, for  
 CP-even (blue) and CP-odd (orange) sneutrinos. Also shown are the FermiLAT limit from dSphs at present (solid black) and as projection for 15 years from now (dashed black).
All points obey the relic density upper limit, for which rescaling, where necessary, has been applied.}
	\label{fig:Scan-FermiLAT}
\end{figure}

When the sneutrino contribute to the observed or a part of DM abundance, its annihilation to SM particles produces an energetic spectrum of SM particles which has chances of being measured in DM indirect detection experiments. 
In this section, we will focus on the photon spectrum, produced as secondaries when sneutrino DM annihilates to SM final states. We will analyse the impact of FermiLAT searches from dwarf spheroidal galaxies (dSphs) and the galactic center in order to constrain and understand the future potential to explore sneutrino DM. 
The annihilation of sneutrinos in  astrophysical objects with DM density $\rho_{\rm DM}$ yields a $\gamma-$ray flux which is given by 
\bea
\frac{d \Phi}{d E_{\gamma}} = \left(\frac{1}{4\pi}\,\frac{\langle \sigma v \rangle}{2 m^2_{\rm DM}}\,\frac{d N_{\gamma}}{d E_{\gamma}}\right)\times\left(\,\int_{\Delta \Omega}\int_{l.o.s.}\rho^2_{\rm DM}\,dl d\Omega'\right)\,,
\label{eq:flux}
\eea
where it  is possible to separate a particle-dependent part, as the cross section $\langle \sigma v \rangle$ and the differential distribution $d N_{\gamma}/d E_{\gamma}$, from the astrophysical term involving the integration of $\rho_{\rm DM}$ over the line-of-sight (l.o.s) and the solid angle $\Delta \Omega$. 
The last term, dubbed J-factor, depends on the particular $\gamma-$ray source where the DM annihilation takes place. The FermiLAT experiment has searched for $\gamma-$rays production with a sensitivity in the energy range from $20$ MeV to $\sim 300$ GeV. Now, dSphs of the Milky Way, which are expected to have a sizable DM content, have a J-factor of $10^{19}$ GeV$^2$ cm$^{-5}$ and a small non-thermal $\gamma-$ray background. These features make their observation particularly suitable in constraining $\langle \sigma v \rangle$ and we challenge the BLSSM sneutrino predicition against the bounds coming from 6 years of observation over 15 dSphs \cite{Ackermann:2015zua}.
Consistently with the result of the previous section that, by far, the main \emph{charged} annihilation channel is represented by $W^+W^-$, we have checked that also the biggest constraint is provided in the same channel\footnote{We notice that, when the DM candidate is not fully responsible for the measured relic density, the cross section has been rescaled by an appropriate factor as shown in \cite{Belanger:2015vwa}.}. In Fig.~\ref{fig:Scan-FermiLAT}, we plot the sneutrino annihilation cross section in the $W^+W^-$ channel. We denote the two populations of sneutrino DM candidates namely, CP-odd and CP-even, with two different colours and compare the thermal cross section prediction  with the existing bounds form dSphs (solid line). We also show the projection from 15 years of observation of 60 dSphs sample. While some CP-odd sneutrino candidates can be tested with  future FermiLAT searches, the constraining power for CP-even candidates is far weaker. Most of the parameter space of this model though remains  safely allowed from existing and also   future searches. It is imperative to note that the constraining power of FermiLAT for sneutrino DM is weaker in our scenarios because of underabundant DM component. Moreover, our scan reveals the existence of a section of the GUT-constrained parameter space
ameanable to  investigation in future searches, here represented by the single point above the dashed line. 
%%%%%%%%%%%%%%%%%%%%%%%%%%%%%%%%%%%%%%%%%%%%%%%%%%%
%%%%%%%%%%%%%%%%%%%%%%%%%%%%%%%%%

In a second attempt to confront our model with the FermiLAT observations, we turn to the galactic center and compute the differential $\gamma$-ray flux due to snuetrino annihilation at the center of the Milky Way. The differential distributions for the gamma spectrum as computed in (\ref{eq:flux})  is itself also a subject of dedicated analyses and experimental searches based on FermiLAT data. 
The flux detected has therefore two components, of signal (SIG) and background (BG),
\bea
\frac{d \Phi_{\gamma}}{d E_{\gamma}} = \frac{d \Phi^{\rm BG}_{\gamma}}{d E_{\gamma}} + \frac{d \Phi^{\rm SIG}_{\gamma}}{d E_{\gamma}}
\label{eq.FluxTot}
\eea 
and we computed the signal flux ($d \Phi^{\rm SIG}_{\gamma}/d E_{\gamma}$) for the case of the sneutrino corresponding to the largest annihilation cross section in our scan.
%%%%%%%%%%%%%%%%%%%%%%%%%%%%%%%%%
\begin{figure}[t]
	\centering
	\includegraphics[width=12cm, height=7.0cm]{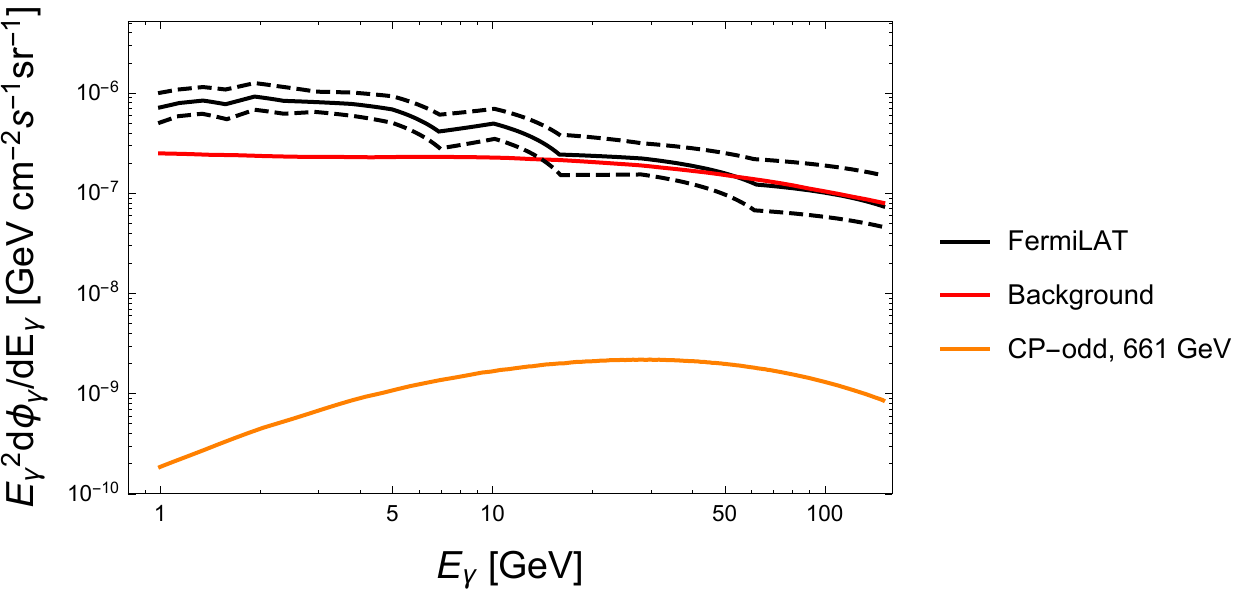}
	\caption{Differential flux of $\gamma$-ray secondary radiation induced by DM DM $\to W^+W^-$ annihilation as a function of the photon energy, with fixed DM mass,
		for our benchmark  CP-odd sneutrino (orange). The corresponding distribution for the background is also given  (red). The FermiLAT present data (with error) are in black.
		The sneutrino point considered is compliant with the relic density constraint taken as an upper limit.}
	\label{fig:Shape-FermiLAT}
\end{figure}
%%%%%%%%%%%%%%%%%%%%%%%%%%%%%%%%%
We notice, as shown in Fig.~\ref{fig:Shape-FermiLAT}, how for our benchmark point of mass of 661 GeV and $\langle\sigma_{WW}\rangle \simeq 7 \times 10^{-25}$ cm$^3$ s$^{-1}$ the signal is far below the large background (given by $ \frac{d \Phi^{\rm BG}_{\gamma}}{d E_{\gamma}}$). Hence,
our  prediction for FermiLAT  is that to a possible detection of a signal in the integrated flux measurement it would not correspond a $\gamma$-ray  spectrum significantly distorted from the background shape, at least not in the current experimental run. However, as the FermiLAT data sample will increase, more and more of the spectrum will be accessible at larger energies, where a characterist signal shape may eventually emerge.

\begin{figure}[h]
	\centering
	\includegraphics[width=13cm, height=7.0cm]{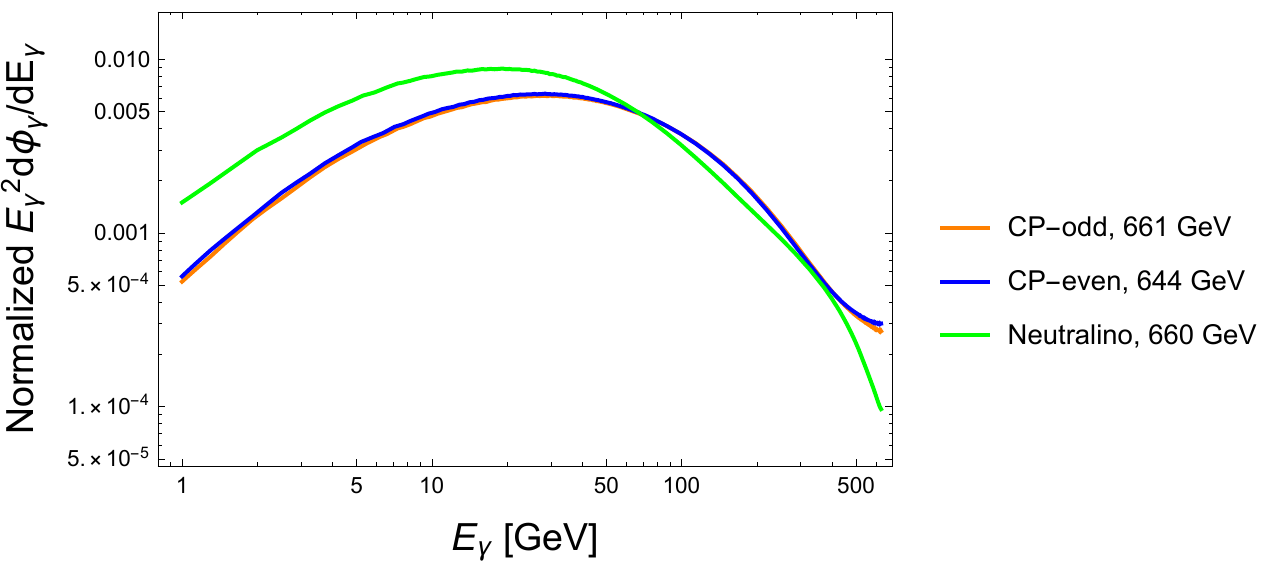} 
	\caption{Differential flux of $\gamma$-ray secondary radiation induced by DM DM $\to W^+W^-$ scatterings as a function of the photon energy, with fixed DM mass,
for  our benchmark CP-even (blue) and CP-odd (orange) sneutrinos. The corresponding distribution for a neutralino is also given for comparison (green). Normalisation is the same for all curves.}
	\label{fig:Shape-DM}
\end{figure}

When this will happen, it will be interesting to understand whether such a shape may enable one to distinguish between a fermionic DM hypothesis and
a CP-even or -odd  one (and possibly between the latter two). With this in mind, 
 we compare the shape of the differential $\gamma$-ray flux from CP-even, CP-odd sneutrino and 
neutralino DM candidates in Fig.~\ref{fig:Shape-DM}. Here, we plot the normalised flux distribution allowing us to make comparison between the three candidates independently of the size of their annihilation cross sections and relic density. The three chosen points have very similar mass, hence also  determining similar end points in the spectrum. While the CP-even and CP-odd sneutrinos have a very similar shape, the neutralino one is very different, this result allowing us to speculate on the possibility of extracting the DM spin  via indirect detection experiments. It should however be noted that a more complete analysis, taking into account various theoretical and experimental uncertainties, must be carried out in order to make a more concrete statement in this direction. Nonetheless, we find this result to be important, as it may actually be testable via data expected to be collected in the years to come.
%%%%%%%%%%%%%%%%%%%%%%%%%%%%%%%%%

\section{LHC Signatures}
\label{sec:LHC}

In this section we discuss the possibility of characterising the sneutrino DM at the LHC by qualitatively describing some of the most interesting signatures provided by the BLSSM. 

Since the LSP sneutrino is mostly RH, it carries no $SU(2)_L$ quantum numbers and hence may only interact with the MSSM-like states via mixing with the LH sneutrinos. This is highly suppressed, being proportional to the very small Dirac Yukawa coupling for the LH neutrinos. 
As such, searches in the neutral or charged Drell-Yan processes, mediated respectively by the SM $Z$ and $W^\pm$ gauge bosons, are hopeless.
In contrast, the largest couplings of the RH sneutrinos are with the typical $(B-L)$ degrees of freedom, among the others, the $Z'$ and heavy bi-leptonic scalars.
In particular, 
as required by CP conservation, the $Z'$ couples to $\snu^{\rm R}$ (CP-even) and $\snu^{\rm I}$ (CP-odd), where one of the two is the LSP and the other can be the Next-to-LSP (NLSP),  
while the heavy CP-even Higgses can couple to two LSPs.  Hence, for the case of {\it direct} DM production at the LHC, one can attempt relying upon   
 $pp \rightarrow Z' \rightarrow \snu_\textrm{LSP} \snu_{\textrm{NLSP}}$, with the decay of the NLSP to the LSP via $\snu_\textrm{NSLP} \rightarrow \snu _{LSP} Z^{(*)}$ providing a di-lepton (plus missing transverse energy) signature through a SM $Z$ boson decay, unlike the heavy Higgs mediated process, which, since the final state is made up by LSP pairs,  is invisible and can only be accessed through  mono-jet, -photon, etc. searches.  In searching for these
direct DM signals,
we have scanned over several benchmark CP-even and CP-odd sneutrino LSPs and used {MadGraph} \cite{Alwall:2014hca} for the computation of the LHC cross sections. 
In detail, we have computed the inclusive cross section for $pp \rightarrow \snu ^{\rm I} _1 \snu ^{\rm R} _i $, where $\snu ^{\rm I} _1$ is the LSP and allowed for the production of any other CP-even sneutrinos ($i=1,\dots, 6$) alongside it. We also have explored the $pp \rightarrow \snu ^{\rm R} _1 \snu ^{\rm I} _i$ channel in which the LSP is represented by the CP-even component of the lightest sneutrino. These cross sections are totally dominated by the $s$-channel exchange of a $Z'$, i.e., $pp \rightarrow Z' \rightarrow \snu ^{\rm I}_1 \snu ^{\rm R}_i, \snu ^{\rm R}_1 \snu ^{\rm I}_i$, and found to be $\sigma \simeq 0.025$ fb at most for both the CP charges of the LSP. It is unsurprising that this cross section is so small, as we are forced to have a heavy $Z'$ to comply with current LHC search limits ($M_{Z'} \gtrsim 4$ TeV). As this cross section is so small, it would be difficult to observe any signal here without a much higher luminosity than at present. 

Another intriguing possibility to search for LSP states though is to do so {\it indirectly}, e.g., 
via slepton $\tilde l$ pair production. The corresponding cross section may lay in the $\sim 0.1$ fb range. 
When the slepton mass is light enough, the $\tilde l \rightarrow W^\pm \snu_\textrm{LSP}$ channel is the only available decay mode despite its width being suppressed by the smallness of the Dirac Yukawa coupling, yiedling a di-lepton signature. Alternatively, if kinematically allowed, one can have $\tilde l \rightarrow \tilde \chi^0 l$ with $\tilde \chi^0 \rightarrow \nu_h \snu_\textrm{LSP}$, where $\nu_h$ is the heavy neutrino. The latter will mainly undergo $\nu_h \rightarrow W^\pm l^\mp$ or $\nu_h \rightarrow Z \nu_l$ decay, thus providing fully or semi-leptonic signatures (again, accompanied by missing transverse energy).  
Other interesting DM signatures may arise from squark pair production for which the cross sections can reach several fb's. In this case, e.g., one can exploit the decay chain $\tilde t \rightarrow \tilde \chi^0 \, t$, which can occur with a BR $\sim$ 80\%  if the $\tilde t$ is the lightest squark, where $\tilde \chi^0 \rightarrow \nu_h \snu_\textrm{LSP}$, as discussed above. Here, one would have a variety of jet plus multi-lepton final states recoiling against 
missing transverse energy.

\section{Conclusions}
\label{sec:conclusion}

The BLSSM provides a preferential DM candidate which is notably different from the MSSM neutralino. The former is a spin-0 boson (specifically, a CP-even or CP-odd sneutrino) and the latter a spin-1/2 fermion (specifically, a neutralino). While in a previous paper we had assessed that  sneutrino DM affords the BLSSM with an amount of parameter space comparatively much larger than  the one
of the MSSM offering  neutralino DM, both compliant with WMAP/Planck and LUX constraints, here, we have shown that signals of sneutrino DM are, on the one hand, just below the 
current sensitivity of FermiLAT and, on the other hand, within reach of it in the next 15 years of foreseen data taking, unlike the neutralino case. Furthermore, we have illustrated that, once a DM signal is established by such an experiment as an excess in the integrated photon flux for some DM mass, there exists scope in establishing the (pseudo)scalar nature of sneutrino DM by studying the differential photon flux in energy, as its shape is notably different from the one pertaining to (fermionic) neutralino DM. However, there exists no possibility in this experiment to separate with differential data the CP-even from the CP-odd sneutrino hypothesis, although their integrated rates are sizely different, with a predominance of relic CP-odd states over CP-even ones.  This phenomenology is enabled by the fact that one of the dominant DM annihilation channels in the case of the BLSSM has charged particles in the final state, notably $W^\pm$ boson pairs, as already  noted in such a previous publication of ours. In fact, it is the copious $\gamma$-ray emission from the charged gauge boson pair that puts FermiLAT in the position of exploring signals of sneutrino DM, unlike   the MSSM, wherein the annihilation channel of neutralino DM into $W^\pm$ pairs is negligible.
 Intriguingly, the favourite BLSSM candidate for DM is also potentially accessible at the LHC over the same time scale, 15 years or so. In fact, 
Run 2 and 3 data from the CERN machine may be able access a series of signatures, involving multi-lepton final states, with and without jets, alongside
the expected missing transverse energy. In fact, also customary mono-jet, -photon, etc. searches may eventually develop sensitivity to the BLSSM candidate for DM.

Altogether, we should like to conclude by mentioning that the DM sector of the BLSSM has very distinctive features with respect to those specific to the prevalent SUSY description, i.e., the MSSM, that can be eventually established in both DM indirect detection experiments and at the LHC. In constrast,
we do not expect (nor we have investigated here) the possibility of differences in case of DM direct searches, as potential BLSSM mediators, a $Z'$ or additional heavy Higgs states, are either too heavy or too weakly coupled to nuclear constituents, respectively, to play any significant role. We therefore advocate more thorough investigations of DM phenomenology in this non-minimal SUSY scenario, which was beyond the scope of our paper.

\section*{Acknowledgements}
SM is supported in part through the NExT Institute. The work of LDR has been supported by the STFC/COFUND Rutherford International Fellowship scheme. The work of CM is supported by the `Angelo Della Riccia' foundation and by the Centre of Excellence project No TK133 `Dark Side of the Universe'. The work of SK is partially supported by the STDF project 13858. SK, SJDK and SM acknowledge support from the grant H2020-MSCA-RISE-2014 n. 645722 (NonMinimalHiggs). SJDK and SM acknowledge support from the STFC Consolidated grant ST/L000296/1.  SuK is supported by the `New Frontiers' program of the Austrian Academy of Sciences and by FWF project number V592-N27.

\newpage

\newpage

\bibliographystyle{JHEP}

\begin{thebibliography}{10}

%\cite{Cao:2017sju}
\bibitem{Cao:2017sju} See, for instance: 
  J.~Cao, L.~Feng, X.~Guo, L.~Shang, F.~Wang, P.~Wu and L.~Zu,
  %``Explaining the DAMPE data with scalar dark matter and gauged $U(1)_{L_e-L_\mu}$ interaction,''
  arXiv:1712.01244 [hep-ph];
  %%CITATION = ARXIV:1712.01244;%%
  %3 citations counted in INSPIRE as of 08 Dec 2017
%\cite{Li:2017tmd}
%\bibitem{Li:2017tmd} 
  T.~Li, N.~Okada and Q.~Shafi,
  %``Scalar dark matter, Type II Seesaw and the DAMPE cosmic ray $e^+ + e^-$ excess,''
  arXiv:1712.00869 [hep-ph];
  %%CITATION = ARXIV:1712.00869;%%
  %3 citations counted in INSPIRE as of 08 Dec 2017
%\cite{Cao:2017ydw}
%\bibitem{Cao:2017ydw} 
  J.~Cao, L.~Feng, X.~Guo, L.~Shang, F.~Wang and P.~Wu,
  %``Scalar dark matter interpretation of the DAMPE data with U(1) gauge interactions,''
  arXiv:1711.11452 [hep-ph];
  %%CITATION = ARXIV:1711.11452;%%
  %15 citations counted in INSPIRE as of 08 Dec 2017
%\cite{Yang:2017zor}
%\bibitem{Yang:2017zor} 
  K.~C.~Yang,
  %``Search for Scalar Dark Matter via Pseudoscalar Portal Interactions: In Light of Galactic Center Gamma-Ray Excess,''
  arXiv:1711.03878 [hep-ph].
  %%CITATION = ARXIV:1711.03878;%%



%\cite{Choubey:2017yyn} 
\bibitem{Choubey:2017yyn} Some recent studies can be listed as: 
  S.~Choubey, S.~Khan, M.~Mitra and S.~Mondal,
  %``Singlet-Triplet Fermionic Dark Matter and LHC Phenomenology,''
  arXiv:1711.08888 [hep-ph];
  %%CITATION = ARXIV:1711.08888;%%
%\cite{Cermeno:2017xwb}
%\bibitem{Cermeno:2017xwb} 
  M.~Cermeño, M.~Á.~Pérez-García and J.~Silk,
  %``Fermionic Light Dark Matter Particles and the New Physics of Neutron Stars,''
  Publ.\ Astron.\ Soc.\ Austral.\  {\bf 34}, e043 (2017)
  %doi:10.1017/pasa.2017.38
  [arXiv:1710.06866 [astro-ph.HE]];
  %%CITATION = %doi:10.1017/pasa.2017.38;%%
%\cite{Chua:2017nxn}
%\bibitem{Chua:2017nxn} 
  C.~K.~Chua and G.~G.~Wong,
  %``Confronting Dirac Fermionic Dark Matter with Recent Data,''
  arXiv:1708.08624 [hep-ph];
  %%CITATION = ARXIV:1708.08624;%%
%\cite{Hwang:2017xmy}
%\bibitem{Hwang:2017xmy} 
  J.~K.~Hwang,
  %``New fermionic dark matters, extended Standard Model and cosmic rays,''
  Mod.\ Phys.\ Lett.\ A {\bf 32}, no. 26, 1730023 (2017);
  %doi:10.1142/S0217732317300233;
  %%CITATION = %doi:10.1142/S0217732317300233;%%
  %1 citations counted in INSPIRE as of 08 Dec 2017
%\cite{Ettefaghi:2017vbh}
%\bibitem{Ettefaghi:2017vbh} 
  M.~Ettefaghi and R.~Moazzemi,
  %``Analyzing of singlet fermionic dark matter via the updated direct detection data,''
  Eur.\ Phys.\ J.\ C {\bf 77}, no. 5, 343 (2017)
  %doi:10.1140/epjc/s10052-017-4894-6
  [arXiv:1705.07571 [hep-ph]];
  %%CITATION = %doi:10.1140/epjc/s10052-017-4894-6;%%
  %1 citations counted in INSPIRE as of 08 Dec 2017
%\cite{Maru:2017otg}
%\bibitem{Maru:2017otg} 
  N.~Maru, T.~Miyaji, N.~Okada and S.~Okada,
  %``Fermion Dark Matter in Gauge-Higgs Unification,''
  JHEP {\bf 1707}, 048 (2017)
  %doi:10.1007/JHEP07(2017)048
  [arXiv:1704.04621 [hep-ph]];
  %%CITATION = %doi:10.1007/JHEP07(2017)048;%%
  %3 citations counted in INSPIRE as of 08 Dec 2017
%\cite{Arbelaez:2017ptu}
%\bibitem{Arbelaez:2017ptu} 
  C.~Arbeláez, M.~Hirsch and D.~Restrepo,
  %``Fermionic triplet dark matter in an $SO(10)$-inspired left right model,''
  Phys.\ Rev.\ D {\bf 95}, no. 9, 095034 (2017)
  %doi:10.1103/PhysRevD.95.095034
  [arXiv:1703.08148 [hep-ph]].
  %%CITATION = %doi:10.1103/PhysRevD.95.095034;%%
  %5 citations counted in INSPIRE as of 08 Dec 2017

%\cite{Arcadi:2017kky}
\bibitem{Arcadi:2017kky} 
  G.~Arcadi, M.~Dutra, P.~Ghosh, M.~Lindner, Y.~Mambrini, M.~Pierre, S.~Profumo and F.~S.~Queiroz,
  %``The Waning of the WIMP? A Review of Models, Searches, and Constraints,''
  arXiv:1703.07364 [hep-ph].
  %%CITATION = ARXIV:1703.07364;%%
  %57 citations counted in INSPIRE as of 08 Dec 2017

%\cite{Akerib:2016lao}
\bibitem{Akerib:2016lao} 
  D.~S.~Akerib {\it et al.} [LUX Collaboration],
  %``Results on the Spin-Dependent Scattering of Weakly Interacting Massive Particles on Nucleons from the Run 3 Data of the LUX Experiment,''
  Phys.\ Rev.\ Lett.\  {\bf 116}, no. 16, 161302 (2016)
  %doi:10.1103/PhysRevLett.116.161302
  [arXiv:1602.03489 [hep-ex]].
  %%CITATION = %doi:10.1103/PhysRevLett.116.161302;%%
  %107 citations counted in INSPIRE as of 08 Dec 2017

%\cite{Aprile:2015uzo}
\bibitem{Aprile:2015uzo} 
  E.~Aprile {\it et al.} [XENON Collaboration],
  %``Physics reach of the XENON1T dark matter experiment,''
  JCAP {\bf 1604}, no. 04, 027 (2016)
  %doi:10.1088/1475-7516/2016/04/027
  [arXiv:1512.07501 [physics.ins-det]].
  %%CITATION = %doi:10.1088/1475-7516/2016/04/027;%%
  %255 citations counted in INSPIRE as of 08 Dec 2017


%\cite{Brink:2005ej}
\bibitem{Brink:2005ej} 
  P.~L.~Brink {\it et al.} [CDMS-II Collaboration],
  %``Beyond the CDMS-II dark matter search: SuperCDMS,''
  eConf C {\bf 041213}, 2529 (2004)
  [astro-ph/0503583].
  %%CITATION = ASTRO-PH/0503583;%%
  %76 citations counted in INSPIRE as of 08 Dec 2017

%\cite{Tanaka:2011uf}
\bibitem{Tanaka:2011uf}
  T.~Tanaka {\it et al.} [Super-Kamiokande Collaboration],
  %``An Indirect Search for WIMPs in the Sun using 3109.6 days of upward-going muons in Super-Kamiokande,''
  Astrophys.\ J.\  {\bf 742}, 78 (2011)
  %doi:10.1088/0004-637X/742/2/78
  [arXiv:1108.3384 [astro-ph.HE]].
  %%CITATION = %doi:10.1088/0004-637X/742/2/78;%%
  %138 citations counted in INSPIRE as of 19 Nov 2016

%\cite{Abbasi:2009uz}
\bibitem{Abbasi:2009uz} 
  R.~Abbasi {\it et al.} [IceCube Collaboration],
  %``Limits on a muon flux from neutralino annihilations in the Sun with the IceCube 22-string detector,''
  Phys.\ Rev.\ Lett.\  {\bf 102}, 201302 (2009)
  %doi:10.1103/PhysRevLett.102.201302
  [arXiv:0902.2460 [astro-ph.CO]].
  %%CITATION = %doi:10.1103/PhysRevLett.102.201302;%%
  %196 citations counted in INSPIRE as of 08 Dec 2017

%\cite{Atwood:2009ez}
\bibitem{Atwood:2009ez} 
  W.~B.~Atwood {\it et al.} [Fermi-LAT Collaboration],
  %``The Large Area Telescope on the Fermi Gamma-ray Space Telescope Mission,''
  Astrophys.\ J.\  {\bf 697}, 1071 (2009)
  %doi:10.1088/0004-637X/697/2/1071
  [arXiv:0902.1089 [astro-ph.IM]].
  %%CITATION = %doi:10.1088/0004-637X/697/2/1071;%%
  %1930 citations counted in INSPIRE as of 08 Dec 2017

%\cite{Abdallah:2016ygi}
\bibitem{Abdallah:2016ygi} 
  H.~Abdallah {\it et al.} [H.E.S.S. Collaboration],
  %``Search for dark matter annihilations towards the inner Galactic halo from 10 years of observations with H.E.S.S,''
  Phys.\ Rev.\ Lett.\  {\bf 117}, no. 11, 111301 (2016)
  %doi:10.1103/PhysRevLett.117.111301
  [arXiv:1607.08142 [astro-ph.HE]].
  %%CITATION = %doi:10.1103/PhysRevLett.117.111301;%%
  %61 citations counted in INSPIRE as of 08 Dec 2017

%\cite{Buchmueller:2017qhf}
\bibitem{Buchmueller:2017qhf} 
  O.~Buchmueller, C.~Doglioni and L.~T.~Wang,
  %``Search for dark matter at colliders,''
  Nature Phys.\  {\bf 13}, no. 3, 217 (2017).
  %doi:10.1038/nphys4054
  %%CITATION = %doi:10.1038/nphys4054;%%
  %1 citations counted in INSPIRE as of 08 Dec 2017


%\cite{Basalaev:2017hni}
\bibitem{Basalaev:2017hni} 
  A.~Basalaev [ATLAS Collaboration],
  %``Search for WIMP dark matter produced in association with a Z boson with the ATLAS detector,''
  EPJ Web Conf.\  {\bf 164}, 08008 (2017).
  %doi:10.1051/epjconf/201716408008
  %%CITATION = %doi:10.1051/epjconf/201716408008;%%

%\cite{Chatrchyan:2012me}
\bibitem{Chatrchyan:2012me} 
  S.~Chatrchyan {\it et al.} [CMS Collaboration],
  %``Search for dark matter and large extra dimensions in monojet events in $pp$ collisions at $\sqrt{s}=7$ TeV,''
  JHEP {\bf 1209}, 094 (2012)
  %doi:10.1007/JHEP09(2012)094
  [arXiv:1206.5663 [hep-ex]].
  %%CITATION = %doi:10.1007/JHEP09(2012)094;%%
  %257 citations counted in INSPIRE as of 08 Dec 2017


%\cite{Hinshaw:2012aka}
\bibitem{Hinshaw:2012aka} 
  G.~Hinshaw {\it et al.} [WMAP Collaboration],
  %``Nine-Year Wilkinson Microwave Anisotropy Probe (WMAP) Observations: Cosmological Parameter Results,''
  Astrophys.\ J.\ Suppl.\  {\bf 208}, 19 (2013)
  %doi:10.1088/0067-0049/208/2/19
  [arXiv:1212.5226 [astro-ph.CO]].
  %%CITATION = %doi:10.1088/0067-0049/208/2/19;%%
  %2735 citations counted in INSPIRE as of 08 Dec 2017

%\cite{Ade:2015xua}
\bibitem{Ade:2015xua} 
  P.~A.~R.~Ade {\it et al.} [Planck Collaboration],
  %``Planck 2015 results. XIII. Cosmological parameters,''
  Astron.\ Astrophys.\  {\bf 594}, A13 (2016)
  %doi:10.1051/0004-6361/201525830
  [arXiv:1502.01589 [astro-ph.CO]].
  %%CITATION = %doi:10.1051/0004-6361/201525830;%%
  %4577 citations counted in INSPIRE as of 08 Dec 2017


%\cite{Falk:1994es}
\bibitem{Falk:1994es} 
  T.~Falk, K.~A.~Olive and M.~Srednicki,
  %``Heavy sneutrinos as dark matter,''
  Phys.\ Lett.\ B {\bf 339}, 248 (1994)
  %doi:10.1016/0370-2693(94)90639-4
  [hep-ph/9409270].
  %%CITATION = %doi:10.1016/0370-2693(94)90639-4;%%
  %254 citations counted in INSPIRE as of 08 Dec 2017

%\cite{Arina:2007tm}
\bibitem{Arina:2007tm} 
  C.~Arina and N.~Fornengo,
  %``Sneutrino cold dark matter, a new analysis: Relic abundance and detection rates,''
  JHEP {\bf 0711}, 029 (2007)
  %doi:10.1088/1126-6708/2007/11/029
  [arXiv:0709.4477 [hep-ph]].
  %%CITATION = %doi:10.1088/1126-6708/2007/11/029;%%
  %126 citations counted in INSPIRE as of 08 Dec 2017

%\cite{Hebbar:2017olk}
\bibitem{Hebbar:2017olk} See, for instance,
  A.~Hebbar, Q.~Shafi and C.~S.~Un,
  %``Light Higgsinos, heavy gluino, and $b-\tau$ quasi-Yukawa unification: Prospects for finding the gluino at the LHC,''
  Phys.\ Rev.\ D {\bf 95}, no. 11, 115026 (2017)
  %doi:10.1103/PhysRevD.95.115026
  [arXiv:1702.05431 [hep-ph]];
  %%CITATION = %doi:10.1103/PhysRevD.95.115026;%%
  %5 citations counted in INSPIRE as of 09 Dec 2017
%\cite{Ahmed:2017ttx}
%\bibitem{Ahmed:2017ttx} 
  W.~Ahmed, L.~Calibbi, T.~Li, S.~Raza, J.~S.~Niu and X.~C.~Wang,
  %``Naturalness and Dark Matter in a Realistic Intersecting D6-Brane Model,''
  arXiv:1711.10225 [hep-ph];
  %%CITATION = ARXIV:1711.10225;%%
%\cite{Chakraborti:2017dpu}
%\bibitem{Chakraborti:2017dpu} 
  M.~Chakraborti, U.~Chattopadhyay and S.~Poddar,
  %``How light a higgsino or a wino dark matter can become in a compressed scenario of MSSM,''
  JHEP {\bf 1709}, 064 (2017)
  %doi:10.1007/JHEP09(2017)064
  [arXiv:1702.03954 [hep-ph]].
  %%CITATION = %doi:10.1007/JHEP09(2017)064;%%
  %6 citations counted in INSPIRE as of 09 Dec 2017












%\cite{DelleRose:2017ukx}
\bibitem{DelleRose:2017ukx} 
  L.~Delle Rose, S.~Khalil, S.~J.~D.~King, C.~Marzo, S.~Moretti and C.~S.~Un,
  %``Naturalness and dark matter in the supersymmetric B-L extension of the standard model,''
  Phys.\ Rev.\ D {\bf 96}, no. 5, 055004 (2017)
  %doi:10.1103/PhysRevD.96.055004
  [arXiv:1702.01808 [hep-ph]].
  %%CITATION = %doi:10.1103/PhysRevD.96.055004;%%
  %6 citations counted in INSPIRE as of 08 Dec 2017


%\cite{Wendell:2010md}
\bibitem{Wendell:2010md} 
  R.~Wendell {\it et al.} [Super-Kamiokande Collaboration],
  %``Atmospheric neutrino oscillation analysis with sub-leading effects in Super-Kamiokande I, II, and III,''
  Phys.\ Rev.\ D {\bf 81}, 092004 (2010)
  %doi:10.1103/PhysRevD.81.092004
  [arXiv:1002.3471 [hep-ex]].
  %%CITATION = %doi:10.1103/PhysRevD.81.092004;%%
  %306 citations counted in INSPIRE as of 09 Dec 2017

%\cite{ATLAS:2017wce}
\bibitem{ATLAS:2017wce} 
  The ATLAS Collaboration, % [ATLAS Collaboration],
  %``Search for new high-mass phenomena in the dilepton final state using 36.1 fb$^{-1}$ of proton-proton collision data at $\sqrt{s} =$ 13 TeV with the ATLAS detector,''
  ATLAS-CONF-2017-027.
  %%CITATION = ATLAS-CONF-2017-027;%%
  %28 citations counted in INSPIRE as of 09 Dec 2017

%\cite{Araz:2017wbp}
\bibitem{Araz:2017wbp} 
  J.~Y.~Araz, G.~Corcella, M.~Frank and B.~Fuks,
  %``Loopholes in $Z^\prime$ searches at the LHC: exploring supersymmetric and leptophobic scenarios,''
  arXiv:1711.06302 [hep-ph].
  %%CITATION = ARXIV:1711.06302;%%
  %2 citations counted in INSPIRE as of 09 Dec 2017


%\cite{Abbas:2007ag}
\bibitem{Abbas:2007ag} 
  M.~Abbas and S.~Khalil,
  %``Neutrino masses, mixing and leptogenesis in TeV scale $B$ - L extension of the standard model,''
  JHEP {\bf 0804}, 056 (2008)
  %doi:10.1088/1126-6708/2008/04/056
  [arXiv:0707.0841 [hep-ph]].
  %%CITATION = %doi:10.1088/1126-6708/2008/04/056;%%
  %30 citations counted in INSPIRE as of 08 Dec 2017


%\cite{OLeary:2011vlq}
\bibitem{OLeary:2011vlq} 
  B.~O'Leary, W.~Porod and F.~Staub,
  %``Mass spectrum of the minimal SUSY B-L model,''
  JHEP {\bf 1205}, 042 (2012)
  %doi:10.1007/JHEP05(2012)042
  [arXiv:1112.4600 [hep-ph]].
  %%CITATION = %doi:10.1007/JHEP05(2012)042;%%
  %55 citations counted in INSPIRE as of 08 Dec 2017



%\cite{Elsayed:2012ec}
\bibitem{Elsayed:2012ec} 
  A.~Elsayed, S.~Khalil, S.~Moretti and A.~Moursy,
  %``Right-handed sneutrino-antisneutrino oscillations in a TeV scale Supersymmetric B-L model,''
  Phys.\ Rev.\ D {\bf 87}, no. 5, 053010 (2013)
  %doi:10.1103/PhysRevD.87.053010
  [arXiv:1211.0644 [hep-ph]].
  %%CITATION = %doi:10.1103/PhysRevD.87.053010;%%
  %8 citations counted in INSPIRE as of 08 Dec 2017

%\cite{Khalil:2011tb}
\bibitem{Khalil:2011tb} 
  S.~Khalil, H.~Okada and T.~Toma,
  %``Right-handed Sneutrino Dark Matter in Supersymmetric B-L Model,''
  JHEP {\bf 1107}, 026 (2011)
  %doi:10.1007/JHEP07(2011)026
  [arXiv:1102.4249 [hep-ph]].
  %%CITATION = %doi:10.1007/JHEP07(2011)026;%%
  %33 citations counted in INSPIRE as of 09 Dec 2017


%\cite{Abdallah:2014fra}
\bibitem{Abdallah:2014fra} 
  W.~Abdallah, S.~Khalil and S.~Moretti,
  %``Double Higgs peak in the minimal SUSY B-L model,''
  Phys.\ Rev.\ D {\bf 91}, no. 1, 014001 (2015)
  %doi:10.1103/PhysRevD.91.014001
  [arXiv:1409.7837 [hep-ph]].
  %%CITATION = %doi:10.1103/PhysRevD.91.014001;%%
  %10 citations counted in INSPIRE as of 09 Dec 2017

%\cite{Lee:2007mt}
\bibitem{Lee:2007mt} 
  H.~S.~Lee, K.~T.~Matchev and S.~Nasri,
  %``Revival of the thermal sneutrino dark matter,''
  Phys.\ Rev.\ D {\bf 76}, 041302 (2007)
  %doi:10.1103/PhysRevD.76.041302
  [hep-ph/0702223 [HEP-PH]].
  %%CITATION = %doi:10.1103/PhysRevD.76.041302;%%
  %64 citations counted in INSPIRE as of 09 Dec 2017





%\cite{Belanger:2006is}
\bibitem{Belanger:2006is} 
  G.~Belanger, F.~Boudjema, A.~Pukhov and A.~Semenov,
  %``MicrOMEGAs 2.0: A Program to calculate the relic density of dark matter in a generic model,''
  Comput.\ Phys.\ Commun.\  {\bf 176}, 367 (2007)
  %doi:10.1016/j.cpc.2006.11.008
  [hep-ph/0607059].
  %%CITATION = %doi:10.1016/j.cpc.2006.11.008;%%
  %579 citations counted in INSPIRE as of 08 Dec 2017

%\cite{Belanger:2013oya}
\bibitem{Belanger:2013oya} 
  G.~Belanger, F.~Boudjema, A.~Pukhov and A.~Semenov,
  %``micrOMEGAs_3: A program for calculating dark matter observables,''
  Comput.\ Phys.\ Commun.\  {\bf 185}, 960 (2014)
  %doi:10.1016/j.cpc.2013.10.016
  [arXiv:1305.0237 [hep-ph]].
  %%CITATION = %doi:10.1016/j.cpc.2013.10.016;%%
  %450 citations counted in INSPIRE as of 08 Dec 2017

			
	%\cite{Bechtle:2013xfa}
	\bibitem{Bechtle:2013xfa} 
	P.~Bechtle, S.~Heinemeyer, O.~Stål, T.~Stefaniak and G.~Weiglein,
	%``$HiggsSignals$: Confronting arbitrary Higgs sectors with measurements at the Tevatron and the LHC,''
	Eur.\ Phys.\ J.\ C {\bf 74}, no. 2, 2711 (2014)
	% %doi:10.1140/epjc/s10052-013-2711-4
	[arXiv:1305.1933 [hep-ph]].
	%%CITATION = %doi:10.1140/epjc/s10052-013-2711-4;%%
	%243 citations counted in INSPIRE as of 10 Dec 2017

	\bibitem{Bechtle:2013wla} 
	 P.~Bechtle, O.~Brein, S.~Heinemeyer, O.~Stål, T.~Stefaniak, G.~Weiglein and K.~E.~Williams,
	%``$\mathsf{HiggsBounds}-4$: Improved Tests of Extended Higgs Sectors against Exclusion Bounds from LEP, the Tevatron and the LHC,''
	Eur.\ Phys.\ J.\ C {\bf 74}, no. 3, 2693 (2014)
	%%doi:10.1140/epjc/s10052-013-2693-2
	[arXiv:1311.0055 [hep-ph]].
	%%CITATION = %%doi:10.1140/epjc/s10052-013-2693-2;%%
	%see \url{http://higgsbounds.hepforge.org}
	


%\cite{Patrignani:2016xqp}
\bibitem{Patrignani:2016xqp} 
  C.~Patrignani {\it et al.} [Particle Data Group],
  %``Review of Particle Physics,''
  Chin.\ Phys.\ C {\bf 40}, no. 10, 100001 (2016).
  %doi:10.1088/1674-1137/40/10/100001
  %%CITATION = %doi:10.1088/1674-1137/40/10/100001;%%
  %2126 citations counted in INSPIRE as of 08 Dec 2017


%\cite{Ackermann:2015zua}
\bibitem{Ackermann:2015zua} 
  M.~Ackermann {\it et al.} [Fermi-LAT Collaboration],
  %``Searching for Dark Matter Annihilation from Milky Way Dwarf Spheroidal Galaxies with Six Years of Fermi Large Area Telescope Data,''
  Phys.\ Rev.\ Lett.\  {\bf 115}, no. 23, 231301 (2015)
  %doi:10.1103/PhysRevLett.115.231301
  [arXiv:1503.02641 [astro-ph.HE]].
  %%CITATION = %doi:10.1103/PhysRevLett.115.231301;%%
  %517 citations counted in INSPIRE as of 08 Dec 2017

%\cite{Belanger:2015vwa}
\bibitem{Belanger:2015vwa} 
  G.~Belanger, D.~Ghosh, R.~Godbole and S.~Kulkarni,
  %``Light stop in the MSSM after LHC Run 1,''
  JHEP {\bf 1509}, 214 (2015)
  %doi:10.1007/JHEP09(2015)214
  [arXiv:1506.00665 [hep-ph]].
  %%CITATION = %doi:10.1007/JHEP09(2015)214;%%
  %27 citations counted in INSPIRE as of 08 Dec 2017

%\cite{Alwall:2014hca}
\bibitem{Alwall:2014hca}
  J.~Alwall {\it et al.},
  %``The automated computation of tree-level and next-to-leading order differential cross sections, and their matching to parton shower simulations,''
  JHEP {\bf 1407} (2014) 079
  %doi:10.1007/JHEP07(2014)079
  [arXiv:1405.0301 [hep-ph]].
  %%CITATION = %doi:10.1007/JHEP07(2014)079;%%
  %2384 citations counted in INSPIRE as of 12 Dec 2017


\end{thebibliography}

\clearpage
\providecommand{\href}[2]{#2}\begingroup\raggedright\endgroup

\end{document}